# Efficient Computation of Slepian Functions for Arbitrary Regions on the Sphere

Alice P. Bates, *Member, IEEE*, Zubair Khalid, *Member, IEEE*, and Rodney A. Kennedy, *Fellow, IEEE*

*Abstract*—In this paper, we develop a new method for the fast and memory-efficient computation of Slepian functions on the sphere. Slepian functions, which arise as the solution of the Slepian concentration problem on the sphere, have desirable properties for applications where measurements are only available within a spatially limited region on the sphere and/or a function is required to be analyzed over the spatially limited region. Slepian functions are currently not easily computed for large band-limits for an arbitrary spatial region due to high computational and large memory storage requirements. For the special case of a polar cap, the symmetry of the region enables the decomposition of the Slepian concentration problem into smaller subproblems and consequently the efficient computation of Slepian functions for large band-limits. By exploiting the efficient computation of Slepian functions for the polar cap region on the sphere, we develop a formulation, supported by a fast algorithm, for the approximate computation of Slepian functions for an arbitrary spatial region to enable the analysis of modern datasets that support large band-limits. For the proposed algorithm, we carry out accuracy analysis of the approximation, computational complexity analysis, and review of memory storage requirements. We illustrate, through numerical experiments, that the proposed method enables faster computation, and has smaller storage requirements, while allowing for sufficiently accurate computation of the Slepian functions.

*Index Terms*—Spatial-spectral concentration problem, Slepian functions, 2-sphere (unit sphere), spherical harmonics.

## I. Introduction

SIGNALS are naturally defined on a sphere in a large number of real-world applications found in various and diverse branches of science and engineering; including medical imaging [1]–[3], cosmology [4]–[6], acoustics [7], [8], geophysics [9], [10], planetary sciences [11], [12], wireless communication [13], [14] and computer graphics [15], [16], to name a few. In these applications, signals and/or data-sets on the sphere are often analyzed in the harmonic domain which is enabled by the spherical harmonic transform which serves as a well-known counterpart of the Fourier transform [17]. Spherical harmonic functions, or spherical harmonics for short, form an orthonormal basis [17] for signals on the sphere. Signals on the sphere can be reconstructed from a finite number of measurements by expansion in the spherical harmonic basis, provided that the spherical harmonic transform can be accurately computed which requires the samples to be taken on a grid (on the whole sphere) defined by sampling schemes [18], [19].

However, it is common for signals to be measured, reconstructed and/or analyzed within a region of the sphere in many fields including medical imaging [20], signal processing [21], [22], geological studies [9], [10], acoustics [7] and cosmological studies [23], [24], to name a few. For example, the samples are unavailable (or unreliable) at the North and South pole for satellite measurements of the Earth's magnetic or gravitational field [25]. Since the data-sets/measurements are defined over the spatially limited region, the use of the globally defined spherical harmonic basis may not be suitable for signal analysis in these applications. Alternatively, Slepian functions, which arise as the solution of the Slepian concentration problem on the sphere [26]–[28] to find the band-limited functions with optimal energy concentration within a spatial region on the sphere, serve as an orthonormal basis of the space formed by band-limited functions, and therefore are well suited for signal analysis [21], [24], [29], [30] and accurate signal reconstruction (over the spatially limited region) [7], [13], [22] in these applications.

Despite being widely applicable, Slepian functions are currently not computed for large band-limits due to the high computational complexity and large memory storage requirements associated with their computation [31]. The *conventional* method for computing Slepian functions for an arbitrary spatial region on the sphere [28], [32] requires the computation of a $L^2 \times L^2$ matrix, where $L$ denotes the band-limit (formally defined in Section II-A) in the spherical harmonic basis, and the subsequent eigenvalue decomposition of this matrix, both of which are computationally intensive. Furthermore, the storage of such a large matrix on a commonly available desktop computer (with limited memory) also becomes infeasible for large $L$. For the special case of a polar cap spatial region, the computational complexity and storage requirements are manageable for large band-limits as the symmetry of the polar cap region enables the decomposition of Slepian concentration problem into subproblems of smaller size, where the largest matrix is $L \times L$ [25], [31]. To the best of our knowledge, Slepian functions have only been computed up to $L = 72$ [33] despite higher band-limit data

Manuscript received August 18, 2016; revised January 10, 2017, April 2, 2017, and April 28, 2017; accepted May 27, 2017. Date of publication June 5, 2017; date of current version June 23, 2017. The associate editor coordinating the review of this manuscript and approving it for publication was Dr. Ian Clarkson. The work of Alice P. Bates was supported by the Australian Research Council's Discovery Projects Funding Scheme (Project no. DP150101011). The work of Rodney A. Kennedy was supported by the Australian Research Council's Discovery Projects Funding Scheme (Project no. DP170101897). *(Corresponding author: Alice P. Bates.)*

The authors are with the Research School of Engineering, The Australian National University, Canberra ACT 0200, Australia (e-mail: alice.bates@anu.edu.au; zubair.khalid@lums.edu.pk; rodney.kennedy@anu.edu.au).

Color versions of one or more of the figures in this paper are available online at http://ieeexplore.ieee.org.

Digital Object Identifier 10.1109/TSP.2017.2712122

being available, with the exception of the special case of the region being a polar cap for which Slepian functions have been constructed up to $L = 500$ [31]. While Slepian functions have been used to spatially and spectrally localize a global spherical harmonic power spectrum with large band-limit, Slepian functions, which in this context are used as data tapers, have a small band-limit [12], [29]. In [34] an iterative algorithm is proposed for the computation of the most concentrated eigenfunction which obtains smaller computational complexity than the conventional method but can only compute the first Slepian function.

With the large band-limits supported by modern data-sets on the sphere, such as the Enhanced Magnetic Model (EMM2015) of magnetic field of Earth with band-limit $L = 720$ [35], it is desirable to be able to compute Slepian functions for large band-limits. In this work, with an aim to make computation of Slepian functions manageable for large band-limits, we address the following questions:

- Can we develop a method for calculating Slepian functions that is more computationally efficient than the conventional method?
- How can we reduce the memory storage requirements to make the computation of Slepian functions manageable on a commonly available desktop computer for large $L$?
- How does the reduction in computational burden and storage requirements impact on the accuracy of Slepian functions?

In addressing these questions, we organize the rest of the paper as follows. We review the necessary mathematical background for signals on the sphere and spherical harmonics in Section II, before presenting the conventional approach to computing Slepian functions on the sphere. The proposed method for computing Slepian functions is then derived in Section III, where we also analyze the properties of the proposed method and develop an algorithm for implementing the proposed method that is computationally and memory efficient. In Section IV, we illustrate the accuracy, computational complexity and storage requirements of the proposed method compared with the conventional method of computing Slepian functions for the example of mainland Australia. Slepian functions for the example of South America are computed using the proposed method in Section V. Concluding remarks are then made in Section VI.

## II. PROBLEM FORMULATION

To clarify the adopted notation, we briefly review the mathematical background for signals defined on the sphere and their spectral domain representation before presenting Slepian functions on the sphere and stating the problem under consideration. The important notation and mathematical symbols adopted in this paper are summarized in Table I.

### A. Mathematical Background

*1) Signals on the Sphere:* A point on the unit sphere $\mathbb{S}^2$ (also known as the 2-sphere or sphere) is given by a unit vector $\hat{\boldsymbol{x}} \equiv \hat{\boldsymbol{x}}(\theta, \phi) \triangleq (\sin\theta\cos\phi, \sin\theta\sin\phi, \cos\theta)' \in \mathbb{R}^3$, where $\theta \in [0, \pi]$ is the colatitude that is measured with respect to the positive $z$-axis and $\phi \in [0, 2\pi)$ is the longitude which is measured with respect to the positive $x$-axis in the $x-y$ plane, and $(\cdot)'$ denotes the vector transpose operation.

TABLE I
IMPORTANT NOTATION AND MATHEMATICAL SYMBOLS
ADOPTED IN THIS PAPER

| Symbol | Notation |
|---|---|
| $\mathbb{S}^2$ | The unit sphere (also known as sphere, 2-sphere). |
| $L^2(\mathbb{S}^2)$ | Hilbert space of square integrable functions on $\mathbb{S}^2$. |
| $Y_\ell^m(\hat{\boldsymbol{x}})$ | Spherical harmonic function of degree $\ell$ and order $m$. |
| $L$ | Band-limit of a function on the sphere. |
| $\mathcal{H}_L$ | The subspace of band-limited signals on $L^2(\mathbb{S}^2)$. |
| $R$ | Arbitrary region on the sphere. |
| $R_\Theta$ | Polar cap of co-latitude radius $\Theta$ centered at the North pole. |
| $R_\Theta(\hat{\boldsymbol{x}}_c)$ | Rotationally symmetric region centered at $\hat{\boldsymbol{x}}_c = (\theta_c, \phi_c)$. |
| $A$ | Area of $R$. |
| $A_\Theta$ | Area of $R_\Theta$ or $R_\Theta(\hat{\boldsymbol{x}}_c)$. |
| $\mathbf{K}$ | $L^2 \times L^2$ matrix containing inner products of $Y_\ell^m(\hat{\boldsymbol{x}})$ on $R$. |
| $\mathbf{C}$ | The matrix $\mathbf{K}$ for $R_\Theta$. |
| $N$ | Sum of eigenvalues of $\mathbf{K}$, $\lceil N \rceil$ is approximately the number of well-concentrated Slepian functions in $R$. |
| $N_\Theta$ | Sum of eigenvalues of $\mathbf{C}$, $\lceil N_\Theta \rceil$ is approximately the number of well-concentrated Slepian functions in $R_\Theta$ or $R_\Theta(\hat{\boldsymbol{x}}_c)$. |
| $\lambda$ | Concentration ratio of Slepian function within a region. |
| $\mathbf{h}_\alpha$ | Eigenvector of $\mathbf{K}$ corresponding to $\lambda_\alpha$, $\alpha = 1, 2, \ldots, L^2$. |
| $h_\alpha(\hat{\boldsymbol{x}})$ | Slepian function for $R$ corresponding to $\lambda_\alpha$, $\alpha = 1, 2, \ldots, L^2$ calculated using the conventional method. |
| $s_\alpha(\hat{\boldsymbol{x}})$ | Slepian functions for $R_\Theta$ corresponding to $\lambda_\alpha$, $\alpha = 1, 2, \ldots, L^2$. |
| $g_\alpha(\hat{\boldsymbol{x}})$ | Slepian functions for $R_\Theta(\hat{\boldsymbol{x}}_c)$ corresponding to $\lambda_\alpha$, $\alpha = 1, 2, \ldots, L^2$. |
| $\mathbf{P}$ | $N_\Theta \times N_\Theta$ matrix containing inner products of rotationally symmetric Slepian functions on $R$. |
| $M$ | The number of points used to calculate the elements of $\mathbf{P}$ via numerical integration. |
| $\tilde{\mathbf{f}}_a$ | Eigenvector of $\mathbf{P}$ corresponding to $\lambda_a$, $a = 1, 2, \ldots, N_\Theta$. |
| $f_a(\hat{\boldsymbol{x}})$ | Slepian function for $R$ corresponding to $\lambda_a$, $a = 1, 2, \ldots, N_\Theta$ calculated using the method proposed in this paper. |

The set of complex-valued square-integrable functions defined on the sphere forms a Hilbert space denoted by $L^2(\mathbb{S}^2)$ equipped with the inner product given by [17]

$$\langle f, h \rangle \triangleq \int_{\mathbb{S}^2} f(\hat{\boldsymbol{x}}) \overline{h(\hat{\boldsymbol{x}})} \, ds(\hat{\boldsymbol{x}}), \qquad (1)$$

for two functions $f$ and $h$ defined on $\mathbb{S}^2$. Here $ds(\hat{\boldsymbol{x}}) = \sin\theta \, d\theta \, d\phi$ is the differential area element on $\mathbb{S}^2$ and $\overline{(\cdot)}$ denotes the complex conjugate. The inner product induces a norm $\|f\| \triangleq \langle f, f \rangle^{1/2}$. We refer the functions with finite energy (finite induced norm) as signals on the sphere.

*2) Spherical Harmonic Domain Representation:* The spherical harmonic functions are the archetype set of basis functions for $L^2(\mathbb{S}^2)$. The spherical harmonic functions (or spherical harmonics for short) $Y_\ell^m(\hat{\boldsymbol{x}})$ for integer degree $\ell \geq 0$ and integer order $|m| \leq \ell$, where $|\cdot|$ denotes the absolute value, are defined as [17], [36]

$$Y_\ell^m(\hat{\boldsymbol{x}}) = Y_\ell^m(\theta, \phi) = N_\ell^m \, P_\ell^m(\cos\theta) e^{im\phi}, \qquad (2)$$

with

$$N_\ell^m = \sqrt{\frac{2\ell+1}{4\pi} \frac{(\ell-m)!}{(\ell+m)!}}, \qquad (3)$$

and $P_\ell^m$ denotes the associated Legendre function of integer degree $\ell$ and integer order $m$ [17]. The spherical harmonics are orthonormal over the sphere with $\langle Y_\ell^m, Y_p^q \rangle = \delta_{\ell,p}\delta_{m,q}$, where $\delta_{m,q}$ is the Kronecker delta function: $\delta_{m,q} = 1$ for $m = q$ and is zero otherwise.

By the completeness of spherical harmonics, we can expand any signal $f \in L^2(\mathbb{S}^2)$ as

$$f(\hat{\boldsymbol{x}}) = \sum_{\ell=0}^{\infty} \sum_{m=-\ell}^{\ell} (f)_\ell^m Y_\ell^m(\hat{\boldsymbol{x}}), \quad (4)$$

where the equality is understood in terms of convergence in the mean [17] and

$$(f)_\ell^m \triangleq \langle f, Y_\ell^m \rangle = \int_{\mathbb{S}^2} f(\hat{\boldsymbol{x}}) \overline{Y_\ell^m(\hat{\boldsymbol{x}})} \, ds(\hat{\boldsymbol{x}}), \quad (5)$$

denotes the spherical harmonic coefficient of degree $\ell$ and order $m$. The signal $f \in L^2(\mathbb{S}^2)$ is defined to be band-limited at degree $L$ if $(f)_\ell^m = 0$ for $\ell \geq L$. The set of band-limited signals forms an $L^2$ dimensional subspace of $L^2(\mathbb{S}^2)$, which is denoted by $\mathcal{H}_L$. We define the column vector $\mathbf{f} = ((f)_0^0, (f)_1^{-1}, (f)_1^0, (f)_1^1, (f)_2^{-2}, \cdots, (f)_{L-1}^{L-1})'$ of size $L^2$ as the spectral domain representation of a band-limited signal $f \in \mathcal{H}_L$.

*3) Rotation on the Sphere:* Rotation of a function on the sphere can be described in terms of the rotation operator $\mathcal{D}(\varphi, \vartheta, \omega)$ which rotates a function by an angle $\omega \in [0, 2\pi)$ around the $z$-axis, followed by an angle $\vartheta \in [0, \pi]$ around the $y$-axis and finally an angle $\varphi \in [0, 2\pi)$ around the $z$-axis, where the axis and rotations follow a right-handed convention [17]. The inverse of the rotation operator is given by $\mathcal{D}(\varphi, \vartheta, \omega)^{-1} = \mathcal{D}(\pi - \omega, \vartheta, \pi - \varphi)$. Rotation of a function on the sphere is realised by inverse rotation of the coordinate system with

$$(\mathcal{D}(\varphi, \vartheta, \omega) f)(\hat{\boldsymbol{x}}) = f(\mathbf{R}^{-1} \hat{\boldsymbol{x}}), \quad (6)$$

where $\mathbf{R}$ is the $3 \times 3$ rotation matrix corresponding to the rotation operator $\mathcal{D}(\varphi, \vartheta, \omega)$ [17].

*4) Regions on the Sphere:* We use $R$ to denote an arbitrary closed region of the sphere with area $A = \int_R ds(\hat{\boldsymbol{x}})$. $R$ can be irregular in shape and does not need to be convex, it can also be a union of unconnected subregions, with $R = R_1 \cup R_2 \cup \ldots$ [17], [28]. A useful region of the sphere is the polar cap region $R_\Theta \triangleq \{\hat{\boldsymbol{x}}(\theta, \phi) \in \mathbb{S}^2 \, | \, 0 \leq \theta \leq \Theta\}$, parameterized by central angle $\Theta$ formed by the boundary of the polar cap with the positive $z$-axis [25].

### B. Slepian Functions on the Sphere

For signals on the sphere, the Slepian concentration problem [37]–[40], to find the band-limited (or space-limited) functions with optimal energy concentration in the spatial (or spectral) domain, has been extensively investigated [17], [25], [27], [28], [30], [41]. In order to maximize the spatial concentration of a band-limited signal $h \in \mathcal{H}_L$ within the spatial region $R \subset \mathbb{S}^2$, we seek to maximize the spatial concentration (energy) ratio $\lambda$ given by [28]

$$\lambda = \frac{\int_R |h(\hat{\boldsymbol{x}})|^2 ds(\hat{\boldsymbol{x}})}{\int_{\mathbb{S}^2} |h(\hat{\boldsymbol{x}})|^2 ds(\hat{\boldsymbol{x}})}, \quad 0 \leq \lambda < 1. \quad (7)$$

The conventional approach to solving this problem is to express it in the spectral domain as

$$\lambda = \frac{\sum_{\ell=0}^{L-1} \sum_{m=-\ell}^{\ell} \sum_{p=0}^{L-1} \sum_{q=-p}^{p} \overline{(h)_\ell^m} (h)_p^q K_{\ell m, pq}}{\sum_{\ell=0}^{L-1} \sum_{m=-\ell}^{\ell} \overline{(h)_\ell^m} (h)_\ell^m}, \quad (8)$$

where

$$K_{\ell m, pq} \triangleq \int_R \overline{Y_\ell^m(\hat{\boldsymbol{x}})} Y_p^q(\hat{\boldsymbol{x}}) ds(\hat{\boldsymbol{x}}). \quad (9)$$

Using the spectral domain version of the concentration ratio (8), the Slepian concentration problem to maximize the concentration ratio $\lambda$ can be solved as an algebraic eigenvalue problem given by

$$\sum_{p=0}^{L-1} \sum_{q=-p}^{p} K_{\ell m, pq} (h)_p^q = \lambda (h)_\ell^m, \quad (10)$$

with matrix formulation

$$\mathbf{K} \mathbf{h} = \lambda \mathbf{h}, \quad (11)$$

where the matrix $\mathbf{K}$ contains elements $K_{\ell m, pq}$ with similar indexing adopted for $\mathbf{h}$ and has dimension $L^2 \times L^2$. The solution of the eigenvalue problem (11) gives $L^2$ eigenvectors $\mathbf{h}_\alpha$, $\alpha = 1, 2, \ldots, L^2$, where the eigenvalue associated with each eigenvector is denoted $\lambda_\alpha$ and eigenvectors are indexed such that $0 \leq \lambda_{L^2} \leq \ldots \leq \lambda_2 \leq \lambda_1 < 1$. Since the eigenvalue problem in (11) is formulated in the spectral domain, each eigenvector represents the spectral domain (spherical harmonic coefficients) of the associated eigenfunction in the spatial domain. The eigenfunctions $h_\alpha(\hat{\boldsymbol{x}})$ are obtained by expanding the eigenvectors $\mathbf{h}_\alpha$ in the spherical harmonic basis using (4). The eigenvalue associated with each eigenvector (or eigenfunction) is a measure of concentration of eigenfunction within the spatial region $R$. Consequently, the eigenfunction $h_1(\hat{\boldsymbol{x}})$ is the most concentrated in $R$, while $h_{L^2}(\hat{\boldsymbol{x}})$ is most concentrated in the complement region $\mathbb{S}^2 \backslash R$.

Since, by definition, $\mathbf{K}$ is Hermitian symmetric and positive semi-definite the eigenvalues are real and non-negative and the eigenvectors can be taken as orthogonal (orthonormal due to normalization in (7) [17], [28]). Such orthogonality, in conjunction with (11), implies

$$\int_{\mathbb{S}^2} h_\alpha(\hat{\boldsymbol{x}}) \overline{h_\beta(\hat{\boldsymbol{x}})} ds(\hat{\boldsymbol{x}}) = \mathbf{h}_\alpha' \overline{\mathbf{h}_\beta} = \delta_{\alpha,\beta}, \quad (12)$$

$$\int_R h_\alpha(\hat{\boldsymbol{x}}) \overline{h_\beta(\hat{\boldsymbol{x}})} ds(\hat{\boldsymbol{x}}) = \mathbf{h}_\alpha' \overline{\mathbf{K} \mathbf{h}_\beta} = \lambda_\beta \mathbf{h}_\alpha' \overline{\mathbf{h}_\beta} = \lambda_\beta \delta_{\alpha,\beta}, \quad (13)$$

that is, the eigenfunctions are orthonormal on the sphere and orthogonal over the region $R$.

*Remark 1 (On the representation of band-limited signals in Slepian basis):* The solution of Slepian concentration problem for a given region $R$ and band-limit $L$ provides $L^2$ orthonormal band-limited eigenfunctions, which span the $L^2$ dimensional subspace $\mathcal{H}_L$ and therefore form a basis, referred to as *Slepian basis* or Slepian functions, for the representation of any signal in $\mathcal{H}_L$.

The number of eigenfunctions that are well-concentrated (with eigenvalue close to 1) in $R$ is approximated[1] by the sum

---

[1] Since the sum of eigenvalues $N$ given in (14) may not be an integer, any reference to $N$ as number is treated as $\lceil N \rceil$ in the rest of the paper to keep the notation succinct. Here $\lceil \cdot \rceil$ denotes the integer ceiling function.

of the eigenvalues $N$, that is,

$$N = \sum_{\alpha=1}^{L^2} \lambda_\alpha = \text{tr}(\mathbf{K}) = \frac{A}{4\pi}L^2, \qquad (14)$$

where $\text{tr}(\cdot)$ denotes the trace of the matrix and $A$ is the area of the region $R$.

*1) Slepian Functions for a Polar Cap:* Here we review the computation of Slepian functions for a polar cap region $R_\Theta$. We use $s \in \mathcal{H}_L$ to denote the functions which we require to be concentrated within the polar cap region. With this consideration, we rewrite the Slepian concentration problem in (11) as

$$\mathbf{C}\mathbf{s} = \lambda \mathbf{s}, \qquad (15)$$

where $\mathbf{C}$ is the matrix $\mathbf{K}$ for a polar cap region and $\mathbf{s}$ is the spectral domain representation of $s$. In order to solve the Slepian concentration problem (10), we are first required to evaluate $K_{\ell m,pq}$, given by the integral over $R$ in (9). For the special case of a polar cap region $R_\Theta$, analytic expressions have been devised in the literature to compute $C_{\ell m,pq}$ [28], [42] in terms of Wigner-$3j$ symbols [17], that is,

$$C_{\ell m,pq} = 2\pi \delta_{m,q} N_\ell^m N_p^m \int_0^\Theta P_\ell^m(\cos\theta) P_p^m(\cos\theta) \sin\theta\, d\theta$$

$$= (-1)^m \frac{\sqrt{(2\ell+1)(2p+1)}}{2} \sum_{n=|\ell-p|}^{\ell+p}$$

$$\times \begin{pmatrix} \ell & n & p \\ 0 & 0 & 0 \end{pmatrix} \begin{pmatrix} \ell & n & p \\ m & 0 & -m \end{pmatrix}$$

$$\times [P_{n-1}(\cos\Theta) - P_{n+1}(\cos\Theta)], \qquad (16)$$

which implies that $C_{\ell m,pq} = 0$ for $m \neq q$ and $C_{\ell m,pq} = C_{\ell(-m),p(-q)}$.

*Remark 2 (On the computation of Slepian functions for polar cap region):* The formulation in (16) implies $C_{\ell m,pq} = 0$ for $m \neq q$ and $C_{\ell m,pq} = C_{\ell(-m),p(-q)}$, which, by appropriate switching of rows and columns of the matrix $\mathbf{C}$, enable us to formulate $\mathbf{C}$ as a block diagonal matrix, where non-zero elements with a fixed order $m$ appear next to each other forming sub-matrices $\mathbf{C}^{(m)}$ of size $(L-m) \times (L-m)$ [25] with

$$\mathbf{C}^{(m)} =$$

$$\begin{pmatrix} C_{mm,mm} & C_{mm,(m+1)m} & \cdots & C_{mm,(L-1)m} \\ C_{(m+1)m,mm} & C_{(m+1)m,(m+1)m} & \cdots & C_{(m+1)m,(L-1)m} \\ \vdots & \vdots & \ddots & \vdots \\ C_{(L-1)m,mm} & C_{(L-1)m,(m+1)m} & \cdots & C_{(L-1)m,(L-1)m} \end{pmatrix}$$

$$(17)$$

for $0 \leq m < L$ and $\mathbf{C}^{(m)} = \mathbf{C}^{(-m)}$. Due to the block diagonal structure of $\mathbf{C}$ for the polar cap region, rather than solving the $L^2 \times L^2$ eigenvalue equation (11), we can solve $L$ smaller problems of size $(L-m) \times (L-m)$, the largest being of size $L \times L$ for $m = 0$, of the form

$$\mathbf{C}^{(m)} \mathbf{s}^{(m)} = \lambda \mathbf{s}^{(m)}, \qquad (18)$$

where $\mathbf{s}^{(m)} = \left((s)_{|m|}^m, (s)_{|m|+1}^m, \cdots, (s)_{L-1}^m\right)'$ contains spherical harmonic coefficients of order $m$. For each eigenvector, the associated Slepian functions can be obtained using (4).

Alternatively, Slepian functions can be obtained directly and efficiently using the method presented in [25], [28]. This method is analytic and so allows for the accurate and fast computation of Slepian functions in a polar cap. The only matrix is a tridiagonal matrix of size $(L-m) \times (L-m)$, the largest being of size $L \times L$ for $m = 0$, that has elements with simple analytical expressions.

The number of Slepian functions that are well-concentrated in the polar cap can be approximated by substituting its area $A_\Theta \triangleq 2\pi(1 - \cos\Theta)$ into (14), giving the sum of eigenvalues of $\mathbf{C}$ for a polar cap region,

$$N_\Theta = \frac{(1 - \cos\Theta)}{2} L^2. \qquad (19)$$

*C. Problem Statement*

Following the formulation of Slepian concentration problem presented above, we summarize below the method, referred to throughout this method as the *conventional method*, for the computation of Slepian functions for a given band-limit $L$ and an arbitrary shaped region $R$:[2]
1) Calculate the $L^2 \times L^2$ matrix $\mathbf{K}$ composed of inner products between spherical harmonic functions (9) via numerical integration of spherical harmonics over $R$.[3]
2) Carry out the eigenvalue decomposition of $\mathbf{K}$ to compute eigenvalues $\lambda_\alpha$ and eigenvectors $\mathbf{h}_\alpha$ for $\alpha = 1, 2, \ldots, L^2$. Eigenvectors represent Slepian functions in the spectral (spherical harmonic) domain.

Computing Slepian functions with large band-limits using this method is infeasible due to the large computational complexity of calculating the $L^4$ elements of $\mathbf{K}$, and subsequently computing the eigenvalues and eigenvectors. At large $L$, the memory required to store $\mathbf{K}$ also becomes too large for a standard desktop computer. For the special case of a polar cap region, following Remark 2, the computational complexity and storage requirements to compute Slepian functions at large band-limits are manageable due to the matrices $\mathbf{C}^{(m)}$ (and the tridiagonal matrix if the eigenfunctions are to be computed directly) being at most of size $L \times L$. In this work, we aim to develop a method of computing Slepian functions for large band-limits and an arbitrary region on the sphere that has manageable computational complexity and storage requirements.

### III. Efficient Computation of Slepian Functions for an Arbitrary Region on the Sphere

As explained in the previous section, the conventional approach, that is computationally expensive and memory inefficient, to compute Slepian functions is to solve the eigenvalue problem in (11), which is obtained by expanding the function $h \in \mathcal{H}_L$ in harmonic space in the equivalent concentration problem formulated in (7). By finding the alternative basis, in which the function $h \in \mathcal{H}_L$ in (7) has sparse representation, we propose a fast method, with less storage (memory) requirements, for the computation of Slepian functions on the sphere for a given band-limit $L$ and region $R \subset \mathbb{S}^2$.

---

[2]This conventional method of computing Slepian functions is implemented in the `SLEPIAN_Alpha` software DOI: 10.5281/zenodo.56825 [43].

[3]Since there does not exist any (exact) quadrature rule to evaluate the integral of the function on the sphere over an arbitrary region, $K_{\ell m,pq}$ given in (9), is computed numerically by employing the approximate quadrature rule [32] for the discretization of integral over arbitrary region $R$.

### A. Slepian Functions for Rotationally Symmetric Region

*Definition 1 (Rotationally Symmetric Region):* We define a rotationally symmetric region centered at $\hat{\boldsymbol{x}}_c = \hat{\boldsymbol{x}}_c(\theta_c, \phi_c)$ enclosing the region $R$[4] as $R_\Theta(\hat{\boldsymbol{x}}_c) = \{\hat{\boldsymbol{x}}(\theta_c, \phi_c) \in \mathbb{S}^2 \,|\, \Delta(\hat{\boldsymbol{x}} \cdot \hat{\boldsymbol{x}}_c) \leq \Theta\}$, where $\Delta(\hat{\boldsymbol{x}} \cdot \hat{\boldsymbol{x}}_c)$ denotes the great circle distance between $\hat{\boldsymbol{x}}$ and $\hat{\boldsymbol{x}}_c$. The region is rotationally symmetric around its center $\hat{\boldsymbol{x}}_c$. We note that $R_\Theta = R_\Theta(\hat{\boldsymbol{x}}_c(0,0))$, that is, a polar cap region is a special case of rotationally symmetric region with $\hat{\boldsymbol{x}}_c = \hat{\boldsymbol{x}}_c(0,0)$ (North pole).

For a given band-limit and rotationally symmetric region $R_\Theta(\hat{\boldsymbol{x}}_c)$, we denote Slepian functions by $g_\alpha(\hat{\boldsymbol{x}})$, $\alpha = 1, 2, \ldots, L^2$. Noting that the polar cap region $R_\Theta$, when rotated around $y$-axis by $\theta_c$ and then by $\phi_c$ around $z$-axis, becomes rotationally symmetric region $R_\Theta(\hat{\boldsymbol{x}}_c)$, we compute Slepian functions for rotationally symmetric region $R_\Theta(\hat{\boldsymbol{x}}_c)$ by first computing Slepian functions $s_\alpha(\hat{\boldsymbol{x}})$, $\alpha = 1, 2, \ldots, L^2$ in the polar cap region $R_\Theta$ followed by the rotation of polar cap Slepian functions as

$$g_\alpha(\hat{\boldsymbol{x}}) = (\mathcal{D}(\phi_c, \theta_c, 0) s_\alpha)(\hat{\boldsymbol{x}}), \quad \alpha = 1, 2, \ldots, L^2. \quad (20)$$

The rotation of $s_\alpha(\hat{\boldsymbol{x}})$ in the spatial domain using the rotation operator $\mathcal{D}(\phi_c, \theta_c, 0)$ in (20) is carried out in spectral (spherical harmonic) domain as a linear transformation given by [17]

$$(g_\alpha)_\ell^m = e^{-im\phi_c} \sum_{m'=-\ell}^{\ell} d_\ell^{m,m'}(\theta_c)(s_\alpha)_\ell^{m'}, \quad (21)$$

where $d_\ell^{m,m'}(\cdot)$ is the Wigner-$d$ function of degree $\ell$ and orders $m, m'$ [17]. We also note that the number of Slepian functions that are well-concentrated in the region $R_\Theta(\hat{\boldsymbol{x}}_c)$ is approximated by $N_\Theta$ given in (19).

### B. Signal Expansion in Slepian Basis

Since Slepian functions within the polar cap region or rotationally symmetric region can be efficiently computed (Remark 2), we can represent/expand any band-limited function $h \in \mathcal{H}_L$, using the Slepian basis designed for a rotationally symmetric region, that is (see Remark 1)

$$h(\hat{\boldsymbol{x}}) = \sum_{\alpha=1}^{L^2} (h)_\alpha g_\alpha(\hat{\boldsymbol{x}}), \quad (22)$$

where

$$(h)_\alpha \triangleq \langle h, g_\alpha \rangle = \int_{\mathbb{S}^2} h(\hat{\boldsymbol{x}}) \overline{g_\alpha(\hat{\boldsymbol{x}})} ds(\hat{\boldsymbol{x}}), \quad (23)$$

denotes the $\alpha$-th Slepian coefficient.

---

[4]If $R$ is a union of unconnected subregions, then the polar cap should enclose all subregions.

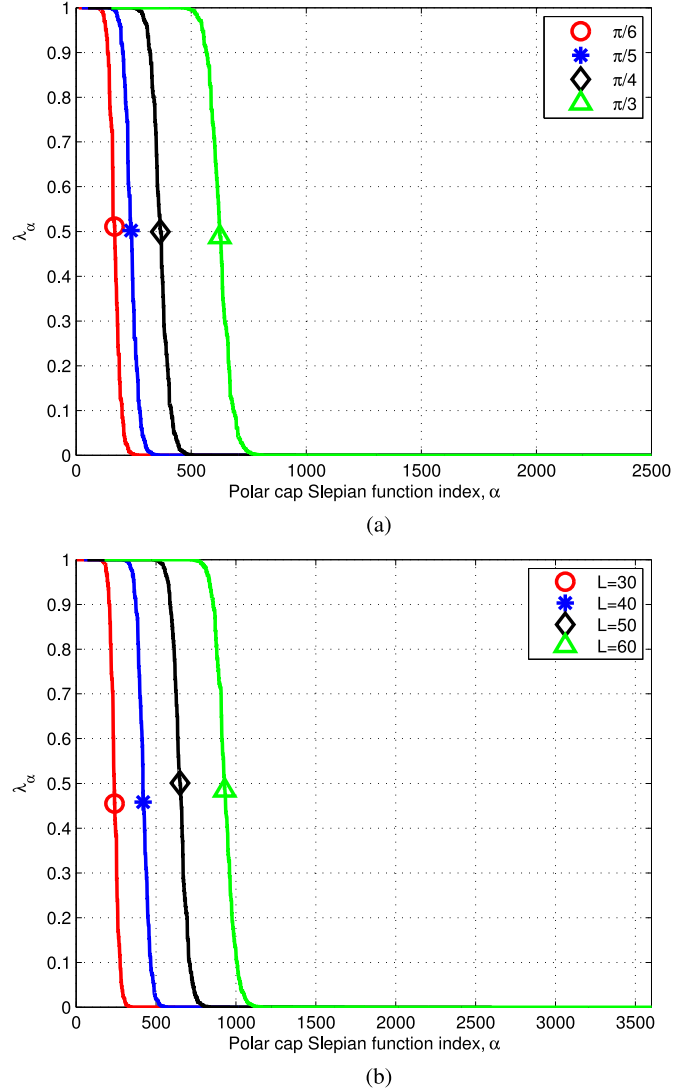

Fig. 1. Eigenvalue spectra for rotationally symmetric regions $R_\Theta(\hat{\boldsymbol{x}}_c)$ centered at the North pole. (a) For a band-limit $L = 50$ and polar cap radii of $\Theta = \pi/6, \pi/5, \pi/4,$ and $\pi/3$. (b) For a polar cap radius of $\Theta = \pi/3$ and for band-limits $L = 30, 40, 50,$ and $60$. $N_\Theta$ is shown by a marker on each spectrum which well-approximates the number of well-concentrated Slepian functions in $R_\Theta(\hat{\boldsymbol{x}}_c)$.

We study the eigenvalue spectrum for rotationally symmetric regions centered at the North pole (polar cap regions) in Fig. 1. Fig. 1(a) shows the eigenvalue spectra for Slepian functions of rotationally symmetric regions centered at the North pole for a band-limit $L = 50$ and polar cap radii of $\Theta = \pi/6, \pi/5, \pi/4,$ and $\pi/3$. Fig. 1(b) shows the eigenvalue spectra for Slepian functions of a rotationally symmetric region with a polar cap radius of $\Theta = \pi/3$ and band-limits $L = 30, 40, 50,$ and $60$. All spectra have a sharp transition from well-concentrated eigenvalues ($\lambda_\alpha \approx 1$) to poorly concentrated eigenvalues ($\lambda_\alpha \approx 0$). This transition takes place at $N_\Theta$, given by (19), as indicated by the markers in Fig. 1.

As only the first $\lceil N_\Theta \rceil$ Slepian functions for the rotationally symmetric region are well-concentrated within the region $R_\Theta(\hat{\boldsymbol{x}}_c)$, the expansion of $h(\hat{\boldsymbol{x}})$ given in (22) can be truncated at[5] $\lceil N_\Theta \rceil$ for $\hat{\boldsymbol{x}} \in R_\Theta(\hat{\boldsymbol{x}}_c)$ as

$$h(\hat{\boldsymbol{x}}) \approx \sum_{\alpha=1}^{N_\Theta} (h)_\alpha g_\alpha(\hat{\boldsymbol{x}}), \quad \hat{\boldsymbol{x}} \in R_\Theta(\hat{\boldsymbol{x}}_c). \quad (24)$$

The error between the Slepian function, given by equation (22), and its approximation, given by (24), within $R_\Theta(\hat{\boldsymbol{x}}_c)$ is

$$\sum_{\alpha=N_\Theta+1}^{L^2} (h)_\alpha g_\alpha(\hat{\boldsymbol{x}}), \quad \hat{\boldsymbol{x}} \in R_\Theta(\hat{\boldsymbol{x}}_c). \quad (25)$$

Due to the small energy concentration $\lambda_\alpha$ within the region $R_\Theta(\hat{\boldsymbol{x}}_c)$ of Slepian functions with $\alpha > N_\Theta$, the error given by (25) is approximately zero.

The quality of the approximation to the Slepian function given in (24) within the spatial region of interest $R$ can be measured by defining the quality measure as a ratio of the energy concentration of the approximate representation to the energy of the exact representation within the spatial region, that is [25], [44],

$$Q(N_\Theta) = \frac{\sum_{\alpha=1}^{N_\Theta} \lambda_\alpha |(h)_\alpha|^2}{\sum_{\alpha=1}^{L^2} \lambda_\alpha |(h)_\alpha|^2}. \quad (26)$$

Later in the paper in Section IV and Section V, we show, through illustration, that the approximate representation given in (24) is of high quality.

### C. Concentration Problem – Formulation in Slepian Basis

We define the truncated expansion given in (24) as

$$f(\hat{\boldsymbol{x}}) \triangleq \sum_{\alpha=1}^{N_\Theta} (f)_\alpha g_\alpha(\hat{\boldsymbol{x}}), \quad (f)_\alpha \triangleq \langle f, g_\alpha \rangle. \quad (27)$$

For an arbitrary shaped region $R \subset R_\Theta(\hat{\boldsymbol{x}}_c)$, we seek to maximize the concentration ratio of $f(\hat{\boldsymbol{x}})$ inside the region $R$ and over the whole sphere, that is,

$$\lambda = \frac{\int_R |f(\hat{\boldsymbol{x}})|^2 ds(\hat{\boldsymbol{x}})}{\int_{\mathbb{S}^2} |f(\hat{\boldsymbol{x}})|^2 ds(\hat{\boldsymbol{x}})} = \text{maximum}, \quad 0 \le \lambda < 1. \quad (28)$$

Using (27), $\lambda$ can be equivalently expressed as

$$\lambda = \frac{\sum_{\alpha=1}^{N_\Theta} \sum_{\beta=1}^{N_\Theta} \overline{(f)_\alpha}(f)_\beta P_{\alpha,\beta}}{\sum_{\alpha=1}^{N_\Theta} \overline{(f)_\alpha}(f)_\alpha}, \quad (29)$$

where

$$P_{\alpha,\beta} \triangleq \int_R \overline{g_\alpha(\hat{\boldsymbol{x}})} g_\beta(\hat{\boldsymbol{x}}) ds(\hat{\boldsymbol{x}}). \quad (30)$$

The problem of maximizing $\lambda$ in (29) can be solved as an algebraic eigenvalue problem of size $N_\Theta$ given by

$$\sum_{\beta=1}^{N_\Theta} P_{\alpha,\beta}(f)_\beta = \lambda(f)_\alpha, \quad (31)$$

with matrix formulation

$$\mathbf{P}\tilde{\mathbf{f}} = \lambda \tilde{\mathbf{f}}, \quad (32)$$

where $\tilde{\mathbf{f}} = ((f)_1, (f)_2, \ldots, (f)_{N_\Theta})'$ and $\mathbf{P}$ is a matrix of size $N_\Theta \times N_\Theta$ with elements given by (30). The solution of the eigenvalue problem in (32) gives $N_\Theta$ orthonormal eigenvectors $\tilde{\mathbf{f}}_a$, $a = 1, 2, \ldots, N_\Theta$ where we have indexed eigenvectors such that eigenvalue $\lambda_a$ associated with the eigenvector $\tilde{\mathbf{f}}_a$ follows $0 \le \lambda_{N_\Theta} \le \ldots \le \lambda_2 \le \lambda_1 < 1$. For each eigenvector $\tilde{\mathbf{f}}_a$, the associated eigenfunction $f_a(\hat{\boldsymbol{x}})$ is obtained using (27).

---

[5]Again, we take any reference to $N_\Theta$ as number as $\lceil N_\Theta \rceil$.

In principle, the Slepian concentration problem in (7) maximizes the concentration of a band-limited function $h \in \mathcal{H}_L$ within the arbitrary spatial region $R$, the solution of which gives $L^2$ band-limited orthonormal eigenfunctions, which serve as an alternative basis, referred to as Slepian basis or functions, for the representation of any band-limited signal. Out of these $L^2$ Slepian functions, $N$ number of Slepian functions are well-concentrated within the spatial region. Consequently, any band-limited function, when expanded in Slepian basis, can be well-approximated within the region using the (first) $N$ concentrated Slepian functions. This is the essence of the concentration problem; it enables sparse representation of a signal concentrated within a region of interest by expansion in the Slepian basis. Conventionally, the concentration problem is formulated as an eigenvalue problem, (11), the solution of which requires eigenvalue decomposition of $L^2 \times L^2$ matrix.

Here we have posed a concentration problem to maximize the concentration of $f(\hat{\boldsymbol{x}})$ within the spatial region $R \subset R_\Theta(\theta_c)$. Since $f(\hat{\boldsymbol{x}}) \approx h(\hat{\boldsymbol{x}})$ for $\hat{\boldsymbol{x}} \in R_\Theta(\theta_c)$, we have $h_\alpha(\hat{\boldsymbol{x}}) \approx f_\alpha(\hat{\boldsymbol{x}})$ for $\alpha = 1, 2, \ldots, N$, that is we have the (approximately) same well-concentrated eigenfunctions of the two concentration problems formulated in (7)–(11) and (28)–(32). However, the latter requires the eigenvalue decomposition of matrix $\mathbf{P}$ of size $N_\Theta \times N_\Theta$ and, therefore, can be solved efficiently and has manageable storage requirements. Consequently, the proposed formulation enables the approximate computation of Slepian functions for a given band-limit $L$ and an arbitrary region $R$. We analyze the accuracy, computational complexity and storage requirements in the next section.

*Remark 3 (On the accurate computation of Slepian Functions for arbitrary regions):* Our proposed method can compute Slepian functions for any arbitrary region of the sphere, which does not have to be well-approximated by a rotationally symmetric region. Slepian functions serve as an orthonormal basis for the whole sphere and an orthogonal basis for the region on which they are defined. Since this is true for any region including the polar cap, the polar cap Slepian functions can be used to represent *any* band-limited function on the sphere (See Remark 1 and (22))). The function can be defined on any region, not just within the polar cap. If the function is concentrated within the polar cap though, it can be represented more efficiently (Slepian functions with a small amount of energy in the region can be discarded, see (24)). Similarly, if the function exists within a region enclosed by the polar cap it can be efficiently represented by the well-concentrated polar cap Slepian functions. Hence, the region does not have to be a polar cap or approximately a polar cap in shape. In this work, the functions in question are Slepian functions in a region of interest that is enclosed by the polar cap.

### D. Properties of Slepian Functions

*1) Orthogonality:* We show that Slepian functions $h_\alpha$ and $f_\alpha$, computed using the conventional method and the proposed

formulation respectively, exhibit the same orthogonality properties for $\alpha = 1, 2, \ldots, N_\Theta$. By definition, the matrix $\mathbf{P}$ is positive semi-definite and Hermitian symmetric, which implies that its eigenvalues are real and non-negative and the eigenvectors are orthogonal. We choose them to be orthonormal, that is,

$$\tilde{\mathbf{f}}'_a \overline{\tilde{\mathbf{f}}_b} = \sum_{\alpha=1}^{N_\Theta} (f_a)_\alpha \overline{(f_b)_\alpha} = \delta_{a,b}, \quad (33)$$

which is equivalent to the orthonormality of associated eigenfunctions $f_a(\hat{\boldsymbol{x}})$ in the spatial domain, that is,

$$\int_{\mathbb{S}^2} f_a(\hat{\boldsymbol{x}}) \overline{f_b(\hat{\boldsymbol{x}})} ds(\hat{\boldsymbol{x}}) = \delta_{a,b}, \quad (34)$$

which is obtained using the expansion of $f(\hat{\boldsymbol{x}})$ given in (27).

In addition to being orthonormal over the whole sphere, the eigenfunctions $f_a(\hat{\boldsymbol{x}})$ are orthogonal over the region $R$, that is,

$$\int_R f_a(\hat{\boldsymbol{x}}) \overline{f_b(\hat{\boldsymbol{x}})} ds(\hat{\boldsymbol{x}}) = \sum_{\alpha=1}^{N_\Theta} (f_a)_\alpha \sum_{\beta=1}^{N_\Theta} \overline{P_{\alpha,\beta} (f_b)_\beta}$$
$$= \tilde{\mathbf{f}}'_a \overline{\mathbf{P} \tilde{\mathbf{f}}_b} = \lambda_b \tilde{\mathbf{f}}'_a \overline{\tilde{\mathbf{f}}_b} = \lambda_b \, \delta_{a,b}, \quad (35)$$

which follows from the formulation of eigenvalue problem in (32) and the orthonormality relation in (33).

*2) Spectral Domain Representation:* Using the definition of $f(\hat{\boldsymbol{x}})$ in (27) and the definition of the spherical harmonic coefficients (5), the spherical harmonic coefficients $(f)_\ell^m$ are given by

$$(f)_\ell^m \triangleq \int_{\mathbb{S}^2} f(\hat{\boldsymbol{x}}) \overline{Y_\ell^m(\hat{\boldsymbol{x}})} \, ds(\hat{\boldsymbol{x}})$$
$$= \int_{\mathbb{S}^2} \sum_{\alpha=1}^{N_\Theta} (f)_\alpha g_\alpha(\hat{\boldsymbol{x}}) \overline{Y_\ell^m(\hat{\boldsymbol{x}})} \, ds(\hat{\boldsymbol{x}})$$
$$= \sum_{\alpha=1}^{N_\Theta} (f)_\alpha \int_{\mathbb{S}^2} g_\alpha(\hat{\boldsymbol{x}}) \overline{Y_\ell^m(\hat{\boldsymbol{x}})} \, ds(\hat{\boldsymbol{x}}) = \sum_{\alpha=1}^{N_\Theta} (f)_\alpha (g_\alpha)_\ell^m. \quad (36)$$

*3) Number of Well-concentrated Eigenfunctions:* With an assumption that the spectrum of eigenvalues has a sharp transition from unity to zero, the number of well-concentrated eigenfunctions in $R$ is approximated by the trace of the matrix $\mathbf{P}$ with

$$N_P = \mathrm{tr}(\mathbf{P}) = \sum_{\alpha=1}^{N_\Theta} \int_R g_\alpha(\hat{\boldsymbol{x}}) \overline{g_\alpha(\hat{\boldsymbol{x}})} ds(\hat{\boldsymbol{x}})$$
$$= \int_R \sum_{\alpha=1}^{N_\Theta} |g_\alpha(\hat{\boldsymbol{x}})|^2 ds(\hat{\boldsymbol{x}}). \quad (37)$$

The sum of Slepian functions over the sphere is independent of the position on the sphere [28], that is,

$$\sum_{\alpha=1}^{L^2} |g_\alpha(\hat{\boldsymbol{x}})|^2 ds(\hat{\boldsymbol{x}}) = \frac{N_\Theta}{A_\Theta}, \quad \hat{\boldsymbol{x}} \in \mathbb{S}^2. \quad (38)$$

Noting that Slepian functions $g_\alpha(\hat{\boldsymbol{x}})$ for rotationally symmetric region have low energy concentration ($\lambda \approx 0$) when $\alpha = N_\Theta + 1, N_\Theta + 2, \ldots, L^2$, we have

$$\sum_{\alpha=1}^{N_\Theta} |g_\alpha(\hat{\boldsymbol{x}})|^2 ds(\hat{\boldsymbol{x}}) \approx \frac{N_\Theta}{A_\Theta}, \quad \forall \hat{\boldsymbol{x}} \in R_\Theta(\hat{\boldsymbol{x}}_c), \quad (39)$$

which allows us to approximate $N_P$ in (37) as

$$N_P \approx \frac{A}{A_\Theta} N_\Theta = \frac{A}{A_\Theta} \frac{A_\Theta L^2}{4\pi} = \frac{A L^2}{4\pi} = N, \quad (40)$$

which indicates both the conventional method and the proposed formulation to solve the concentration problem yield approximately the same number of well-concentrated eigenfunctions.

### E. Efficient Computation of Slepian Functions for Arbitrary Region – Algorithm

Based on the formulation presented in Section III, we here present an algorithm to compute Slepian functions for a given band-limit $L$ and arbitrary region $R$. To enable the computation of Slepian functions for large band-limits, the algorithm, consisting of following steps, is designed to minimize the computation time and storage requirements:

1) Find $\Theta$ and $\hat{\boldsymbol{x}}_c$ for a rotationally symmetric region $R_\Theta(\hat{\boldsymbol{x}}_c)$ of smallest area enclosing the region $R$.
2) Rotate $R$ and $R_\Theta(\hat{\boldsymbol{x}}_c)$ to the North pole, first by $\pi - \phi_c$ around $z$-axis, and then by $\theta_c$ around $y$-axis. The rotationally symmetric region rotated to the North pole becomes a polar cap region $R_\Theta$. We use $\tilde{R}$ to denote the region $R$ rotated to the North pole.
3) Compute Slepian functions concentrated in $R_\Theta$ as $s_\alpha(\hat{\boldsymbol{x}})$, $\alpha = 1, 2, \ldots, N_\Theta$ for each order $-L < m < L$:
   a) Find the spherical harmonic coefficients of the polar cap Slepian functions of order $m$ $\mathbf{s}^{(m)}$, by solving (18).
   b) Discard any polar cap Slepian functions that are not well-concentrated in the region.
   c) Evaluate the remaining well-concentrated polar cap Slepian functions in the spatial domain $s(\hat{\boldsymbol{x}})$ by expansion in the spherical harmonic basis (4).
4) Calculate the matrix $\mathbf{P}$, whose elements are given by

$$P_{\alpha,\beta} \triangleq \int_R \overline{g_\alpha(\hat{\boldsymbol{x}})} g_\beta(\hat{\boldsymbol{x}}) ds(\hat{\boldsymbol{x}})$$
$$= \int_{\tilde{R}} \overline{s_\alpha(\hat{\boldsymbol{x}})} s_\beta(\hat{\boldsymbol{x}}) ds(\hat{\boldsymbol{x}}), \quad (41)$$

by numerically integrating the polar cap Slepian functions $s_\alpha(\hat{\boldsymbol{x}})$, $\alpha = 1, 2, \ldots, N_\Theta$ over the region $\tilde{R}$ using the samples stored in Step 3c).
5) Find the eigenvalues and eigenvectors of $\mathbf{P}$.
6) To obtain Slepian functions for $R$, $f_a(\hat{\boldsymbol{x}})$, $a = 1, 2, \ldots, N_\Theta$, (note that for $R_\Theta$, $g_\alpha(\hat{\boldsymbol{x}}) = s_\alpha(\hat{\boldsymbol{x}})$, hence equations (36) and (27) apply for $s_\alpha(\hat{\boldsymbol{x}})$) either:
   a) Calculate the spherical harmonic coefficients of the eigenvectors of $\mathbf{P}$ $(s)_\ell^m$ using (36), then rotate the eigenvectors to the region's original location using (21) and finally expand in the spherical harmonic basis using (4).

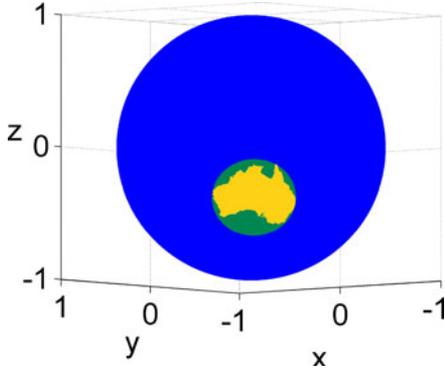

Fig. 2. Mainland Australia on the sphere surrounded by the rotationally symmetric region $R_\Theta(\hat{\boldsymbol{x}}_c)$ shown in green.

  b) Expand the eigenvectors for the region at the North pole in the truncated polar cap basis (27) to get Slepian functions for the region at the North pole before rotating the functions to the region using (6).

We here expand on some of these steps. In Step 1), we determine $\Theta$ and $\hat{\boldsymbol{x}}_c$ for a rotationally symmetric region $R_\Theta(\hat{\boldsymbol{x}}_c)$ enclosing the region $R$ as follows. If $C$ denotes the boundary of the region, we first numerically find the two points $\hat{\boldsymbol{y}}_1, \hat{\boldsymbol{y}}_2 \in C \subset R \subset \mathbb{S}^2$ for which the spherical distance $\Delta(\hat{\boldsymbol{y}}_1, \hat{\boldsymbol{y}}_2) = \cos^{-1}(\hat{\boldsymbol{y}}_1 \cdot \hat{\boldsymbol{y}}_2)$ [17] between them is maximum. This is performed using a search over all pairs of boundary points which has computational complexity $O(N^2)$. We consider the number of points $N$ to be relatively low so that this does not effect the overall computational complexity of the algorithm. For regions with a large number of boundary points, alternative methods for finding the enclosing polar cap, such as first finding the convex hull of the region with complexity $O(N \log N)$, can be investigated. Then, we determine $\Theta$ as

$$\Theta = \frac{\Delta(\hat{\boldsymbol{y}}_1, \hat{\boldsymbol{y}}_2)}{2}, \quad (42)$$

and $\hat{\boldsymbol{x}}_c$ as the center point of the smaller arc of the great circle passing through $\hat{\boldsymbol{y}}_1$ and $\hat{\boldsymbol{y}}_2$. For regions where $\Delta(\hat{\boldsymbol{y}}_1, \hat{\boldsymbol{y}}_2) > \pi$, that is the region $R$ extends onto both hemispheres, a modification is needed. In this case points $\hat{\boldsymbol{y}}_1, \hat{\boldsymbol{y}}_2 \in C \subset R \subset \mathbb{S}^2$ need to be found so that the spherical distance is minimized and $\Theta = \pi - \frac{\Delta(\hat{\boldsymbol{y}}_1, \hat{\boldsymbol{y}}_2)}{2}$. As an illustration, the rotationally symmetric region $R_\Theta(\hat{\boldsymbol{x}}_c)$ enclosing mainland Australia region $R$ is shown in Fig. 2.

In Step 3), the polar cap Slepian functions are computed one order $m$ at a time and only the well-concentrated ones are stored to reduce storage requirements from $L^2 \times L^2$ to $N_\Theta \times M$, where $M$ is the number of points that the polar cap Slepian functions are evaluated at for subsequent numerical integration. In Step 4), the inner products between the polar cap Slepian functions are calculated using numerical integration. We use the trapezium rule on an equiangular grid with a resolution parameter that is used to set the number of points used in the integration. The resolution can be increased to allow for greater accuracy or decreased to reduce computation time and storage requirements. Since the equiangular sampling has dense sampling around the poles ($\theta = 0$ or $\theta = \pi/2$), we note that the evaluation of the integral is more accurate, for the same resolution parameter, if the region is closer to poles than the equator ($\theta = \pi/2$).

*F. Efficiency Analysis - Computation Time and Memory*

We here analyze the efficiency in terms of computation time and memory requirements of our proposed algorithm for computing Slepian functions within an arbitrary region on the sphere presented in Section III-E.

The proposed algorithm computes the $N_\Theta \times N_\Theta$ matrix $\mathbf{P}$ by carrying out the inner products of polar cap Slepian functions using numerical integration using $M = L^2$ points, resulting in computational complexity $O(L^2 N_\Theta^{\,2}) = 0.25(1 - \cos\Theta)^2 O(L^6)$, using (19). Note that the number of points $M$ can be decreased to reduce computation time but this will decrease the accuracy of computation. The computational complexity for eigenvalue decomposition of $n \times n$ matrix is $O(n^3)$ [31], hence the eigenvalue decomposition of $\mathbf{P}$ has complexity $O(N_\Theta^{\,3}) = 0.125(1 - \cos\Theta)^3 O(L^6)$. For the conventional method of computing Slepian functions (reviewed in Section II-B), the matrix $\mathbf{K}$ is size $L^2 \times L^2$, integrating a function with band-limit $L$ requires at least $L^2$ samples [18], [45], hence computation of the $L^4$ elements of $\mathbf{K}$ has computational complexity $O(L^6)$. The eigenvalue decomposition of $\mathbf{K}$ is also $O(L^6)$. Hence, the computational complexity of the proposed method is $O(L^6)$, like the conventional, method but with prefactor $0.25(1 - \cos\Theta)^2$ for the matrix computation and $0.125(1 - \cos\Theta)^3$ for the eigenvalue decomposition. It is noted that the parallel computing capability can be used to reduce the computation time of both the proposed and conventional methods of computing the matrix [10].

For applications, where Slepian functions are required to be computed for large band-limits and spatial regions with enclosing rotationally symmetric regions with small area, we expect that the proposed method offers significant reduction in the computation time and memory storage requirement since the reduction in both the memory requirements and the computation time is proportional to the area of the rotationally symmetric region enclosing the spatial region of interest

## IV. ILLUSTRATION - AUSTRALIA

In this section, we evaluate the proposed algorithm, presented in Section III-E, for computing Slepian functions within an arbitrary region on the sphere in terms of the numerical accuracy of Slepian functions and eigenvalues, and the computational complexity and storage requirements. In order to carry out the analysis, we compute Slepian functions for the example of mainland Australia as the region $R$.

As discussed in Section III-E, the resolution of the equiangular grid for numerically integrating the polar cap Slepian functions can be altered to change the accuracy of integration, and the computational and storage requirements. For analysing the proposed algorithm, we set the resolution of equiangular grid so $M \approx L^2$ samples are used for numerical integration.

*A. Numerical Accuracy Analysis - Spatial Domain*

We here compare the accuracy of computing Slepian functions using the proposed method $f_a(\hat{\boldsymbol{x}})$ compared with the conventional method $h_a(\hat{\boldsymbol{x}})$. For a given band-limit $L$, expansion of the band-limited functions using all $L^2$ Slepian functions designed for a rotationally symmetric region, is equivalent to the expansion of a signal in the spherical harmonic basis. Since Slepian functions $g_\alpha(\hat{\boldsymbol{x}})$, $\alpha = N_\Theta + 1, N_\Theta + 2, \ldots, L^2$, although negligible, have some energy in the spatial region $R$, the truncation of the representation in Slepian basis at $N_\Theta$, given in (24), results in an approximation error as we will show in this section. It must be noted neither the proposed nor the conventional method for the computation of Slepian functions is exact due to the need to numerically integrate the spherical harmonics $Y_\ell^m(\hat{\boldsymbol{x}})$ for all degrees $\ell \geq 0$ and orders $|m| \leq \ell$ over $R$ (9) in the conventional method and the polar cap Slepian functions $s_\alpha(\hat{\boldsymbol{x}})$, $\alpha = 1, 2, \ldots, N_\Theta$ over $\tilde{R}$ (41) for the proposed method.

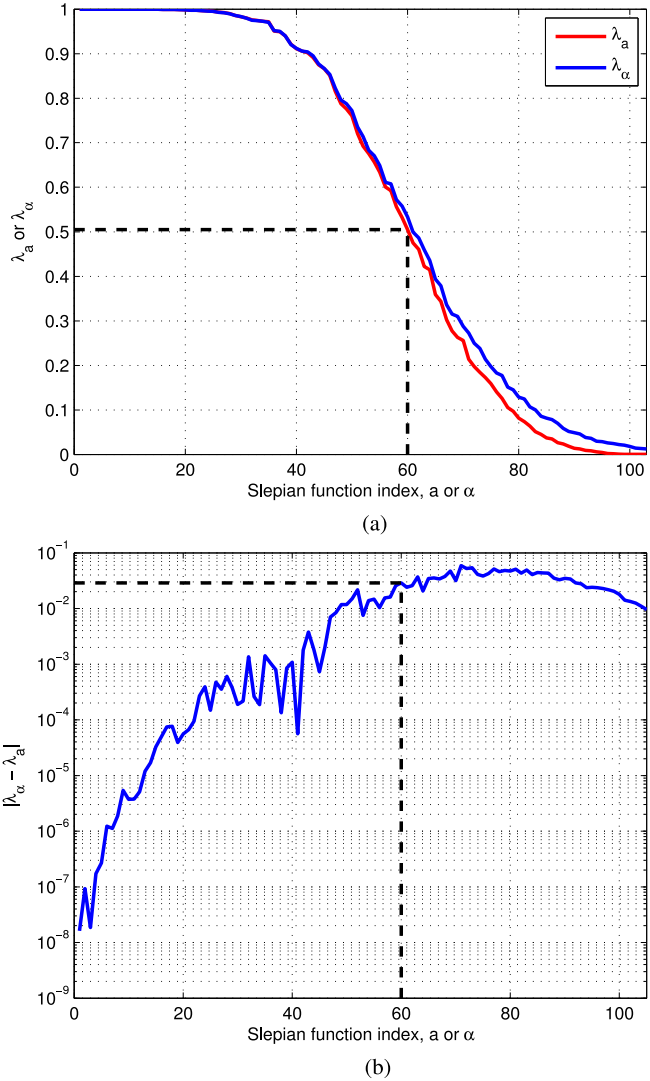

Fig. 3. (a) Eigenvalue spectrum for Slepian functions concentrated in Australia band-limited at $L = 64$ ($N_\Theta = 103$) obtained using the proposed $\lambda_a$, $a = 1, 2, \ldots, N_\Theta$ and conventional $\lambda_\alpha$, $\alpha = 1, 2, \ldots, N_\Theta$ methods. (b) Absolute difference in eigenvalues $|\lambda_\alpha - \lambda_a|$. Dashed back line shows approximate number of well-concentrated Slepian functions $N$.

We compare the numerical accuracy of the two methods for the mainland Australia region $R$ and band-limit $L = 64$. Fig. 3(a) shows the eigenvalue spectrum obtained using the conventional $\lambda_\alpha$, $\alpha = 1, 2, \ldots, N_\Theta$ and proposed method $\lambda_a$, $a = 1, 2, \ldots, N_\Theta$. The absolute difference in the eigenvalues computed using the conventional method and the proposed method $|\lambda_\alpha - \lambda_a|$ is plotted in Fig. 3(b), where it can be observed that the spectra obtained by both methods are similar with the difference in corresponding eigenvalues of the proposed method and conventional method being on the order of $10^{-2}$ or less. Furthermore, the difference is smaller for the most concentrated eigenfunctions in $R$ and grows with the decrease in the spatial concentration of eigenfunctions. This is of significant importance in many applications that only use the well-concentrated Slepian functions for signal analysis [25], [46]. The number of well-concentrated Slepian functions in Australia is approximated by the trace of the matrix $\mathbf{K}$ for the conventional method (14) and by the trace of the matrix $\mathbf{P}$ for the proposed method (37), which both round to $N = 63$, as indicated by the black dashed line in Fig. 3.

We plot the ten most concentrated Slepian functions over Australia shown in Fig. 4 using the conventional and the proposed method, where the similarity in the shape of Slepian functions can be observed. We note that we have plotted the real Slepian functions in Fig. 4; the real Slepian functions can be computed from the complex Slepian functions using the relationship between their real and complex spherical harmonic coefficients [17]. The difference in decibels between the Slepian functions obtained using the conventional and the proposed method, $10\log_{10}|h_\alpha - f_a|$, for the ten most concentrated Slepian functions is shown in Fig. 5. The maximum difference observed in Fig. 5 is smaller than 0 dB showing that the proposed method allows for accurate computation of the Slepian functions.

### B. Numerical Accuracy Analysis - Spectral Domain

To further quantify the difference in Slepian functions obtained using the two methods, we compute the quality factor $Q(N_\Theta)$, given by (26), for each Slepian function band-limited at $L = 64$ and concentrated in mainland Australia computed using the proposed method $f_a$ and plot this in Fig. 6. The quality factor is high for all Slepian functions but particularly for Slepian functions well-concentrated within $R$, with a quality factor of 95% or higher. This shows that the approximation, given in (24), used by the proposed method is highly accurate. The number of well-concentrated eigenfunctions is indicated by $N$, shown by the black dashed line in Fig. 6.

We also calculate the mean difference $E_a$ in the spherical harmonic coefficients of Slepian functions computed using the proposed $(f)_\ell^m$ and conventional $(h)_\ell^m$ methods with

$$E_a \triangleq \frac{1}{L^2} \sum_{\ell=0}^{L-1} \sum_{m=-\ell}^{\ell} |(f_a)_\ell^m - (h_a)_\ell^m|, \qquad (43)$$

this is shown in Fig. 7 for $a = 1, 2, 5$ and $6$, the first, second, fifth and sixth most concentrated Slepian functions in Australia for band-limits $L = 20, 40, 60, 80$ and $100$. $E_a$ is on the order of $10^{-3}$ or smaller, indicating that the proposed method allows the accurate computation of Slepian functions.

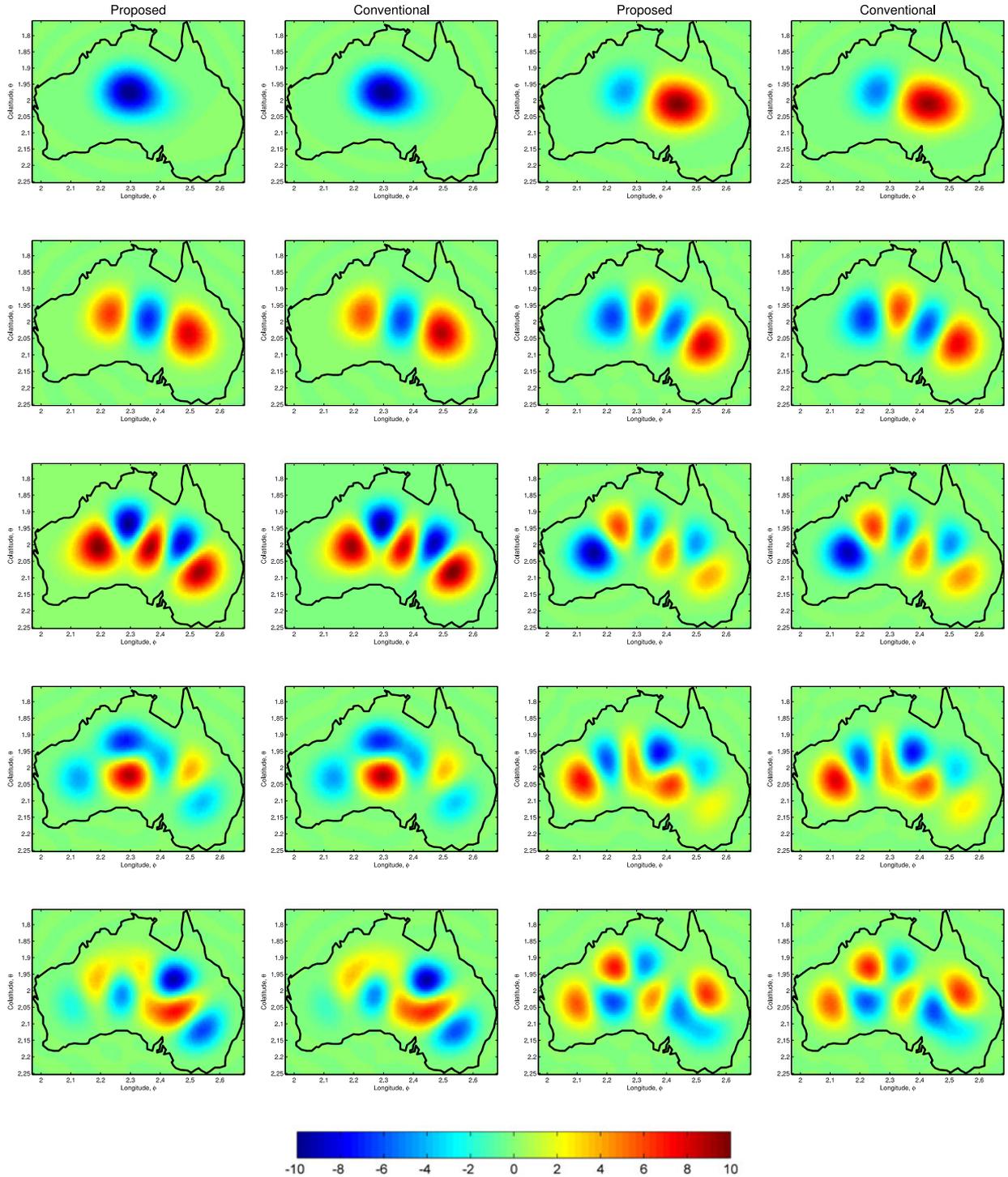

Fig. 4. Slepian functions $f_a(\hat{\boldsymbol{x}})$, $a = 1, 2, \ldots, 10$ and $h_\alpha(\hat{\boldsymbol{x}})$, $\alpha = 1, 2, \ldots, 10$ most concentrated in Australia with band-limit $L = 64$. The ordering of concentration is left to right, top to bottom with Slepian functions obtained using the proposed method $f_a(\hat{\boldsymbol{x}})$ in the first and third columns and Slepian functions obtained using the conventional method $h_\alpha(\hat{\boldsymbol{x}})$ in the second and fourth columns.

## C. Efficiency Analysis

We here analyze the efficiency in terms of the computational complexity and memory required to compute Slepian functions for the proposed method, as discussed in Section III-F for the example of mainland Australia.

Fig. 8 shows the computation time for calculating the matrices $\mathbf{P}$ and $\mathbf{K}$, and subsequently performing their eigenvalue decomposition using the proposed and conventional methods respectively using MATLAB running on a machine equipped with 3.4 GHz Intel Core i7 processor and 8 GB of RAM for main-land Australia. As can be seen in Fig. 8 the computation time of the proposed method is much faster than the conventional method, around two orders of magnitude. The smaller dimension of the **P** matrix compared with **K** results in a faster matrix computation time and faster eigenvalue decomposition.

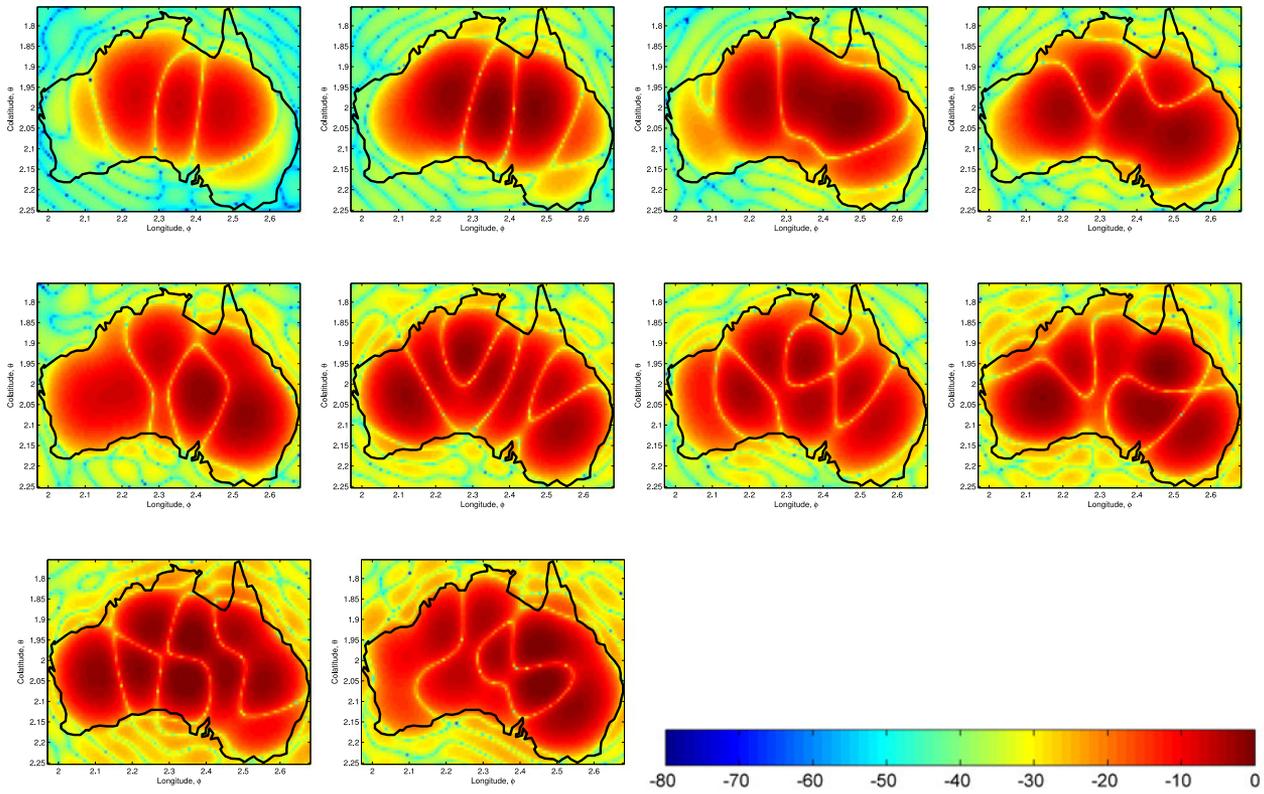

Fig. 5. The difference in decibels between Slepian functions computed using the conventional and the proposed method, $10\log_{10}|h_\alpha - f_a|$, for the ten most concentrated in Australia with band-limit $L = 64$. The ordering of concentration is left to right, top to bottom.

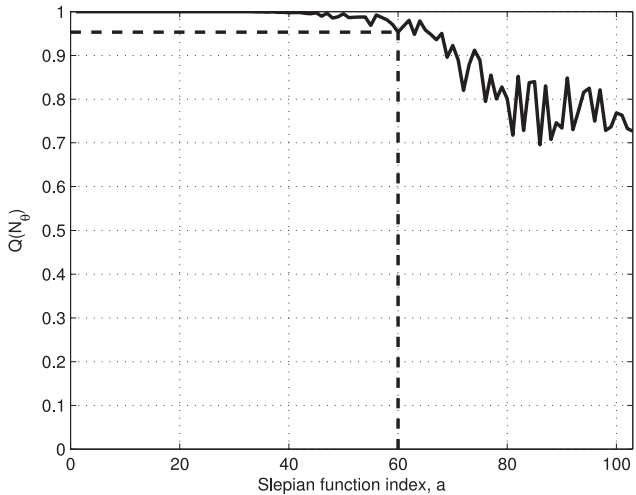

Fig. 6. Quality factor $Q(N_\Theta)$, given by (26), for Slepian functions computed using the proposed method $f_a$. Dashed back line shows approximate number of well-concentrated Slepian functions $N$.

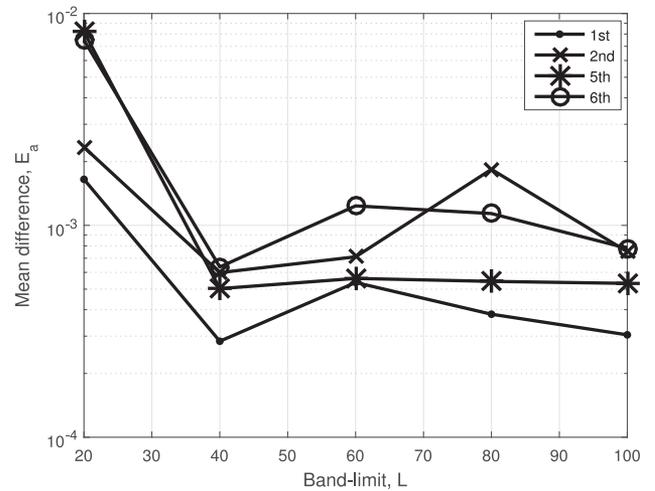

Fig. 7. Mean difference $E_a$ (43) for $a = 1, 2, 5$ and $6$, the 1st, 2nd, 5th and 6th most concentrated Slepian functions in Australia for $L = 20, 40, 60, 80$ and $100$.

The memory required to store matrix **P** is $0.25(1 - \cos\Theta)^2$ times smaller compared with storing the $L^2 \times L^2$ matrix **K**. The commonly available desktop machine used in our analysis has $1.302 \times 10^{10}$ bytes for array storage in MATLAB. As MATLAB's type double requires 64 bits, the maximum band-limit that the matrix **K** can be stored for is $L = 200$. In practice, as matrices other than **K** need to be stored, it is only possible to compute Slepian functions using the conventional method for less than $L = 100$.

The maximum band-limit which the matrix **P** can be stored depends on the region $R$, or more specifically on the area of the rotationally symmetric region enclosing $R$. For example, the rotationally symmetric region surrounding Australia is $2.5\%$ of the area of the sphere hence, the matrix $\mathbf{P}$ is of size $0.000625L^4$ and the maximum band-limit that $\mathbf{P}$ can be stored for is $L = 1270$ for $1.302 \times 10^{10}$ bytes of array storage. In our proposed algorithm presented in Section III-E, Step 3) requires a $N_\Theta \times M$ matrix to store the $N_\Theta$ well-concentrated polar cap Slepian functions evaluated at $M$ points. For $M = L^2$ points used Section IV, the maximum band-limit that this matrix can be stored for on our desktop computer is $L = 505$; the maximum band-limit could be increased by decreasing $M$. We have managed to compute Slepian functions in Australia using the proposed method and $M = L^2$ for band-limits up to $L = 320$.

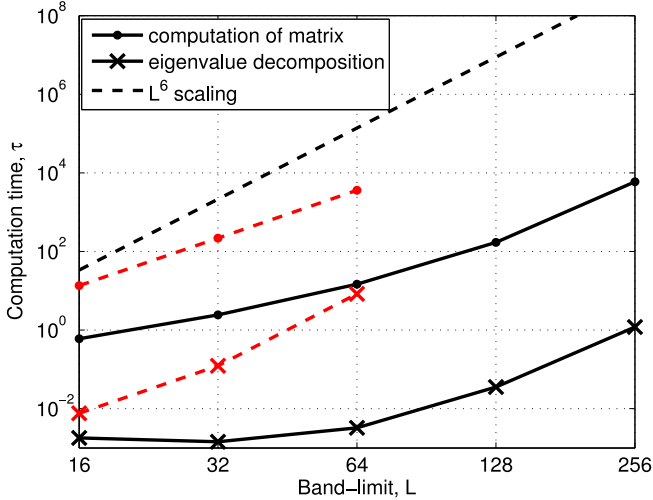

Fig. 8. The computation time $\tau$ in seconds to compute the matrices $\mathbf{P}$ and $\mathbf{K}$ and subsequently perform eigenvalue decomposition using the proposed, shown by the solid black lines, and conventional, shown by the red dashed lines, methods respectively to compute Slepian functions concentrated in Australia for $L = 16, 32, 64, 128$ and $256$.

## V. ILLUSTRATION - SOUTH AMERICA

Our proposed method can compute Slepian functions for any arbitrary region of the sphere, the region does not have to be well-approximated by a rotationally symmetric region (Remark 3). We have included Slepian functions band-limited at $L = 32$ for South America, which is less similar to a rotationally symmetric region than Australia, as another example. Fig. 9(a) shows the eigenvalue spectrum obtained using the conventional $\lambda_\alpha$, $\alpha = 1, 2, \ldots, N_\Theta$ and proposed method $\lambda_a$, $a = 1, 2, \ldots, N_\Theta$. The absolute difference in the eigenvalues computed using the conventional method and the proposed method $|\lambda_\alpha - \lambda_a|$ is plotted in Fig. 9(b), where it can be observed that the spectra are similar with the difference in corresponding eigenvalues of the proposed method and conventional method being on the order of $10^{-2}$ or less, as was the case for the example of Slepian functions with $L = 64$ defined on mainland Australia in Fig. 3.

We plot the ten most concentrated Slepian functions over South America shown in Fig. 10 using the conventional and the proposed method, where the similarity in the shape of Slepian functions can be observed. The difference in decibels between

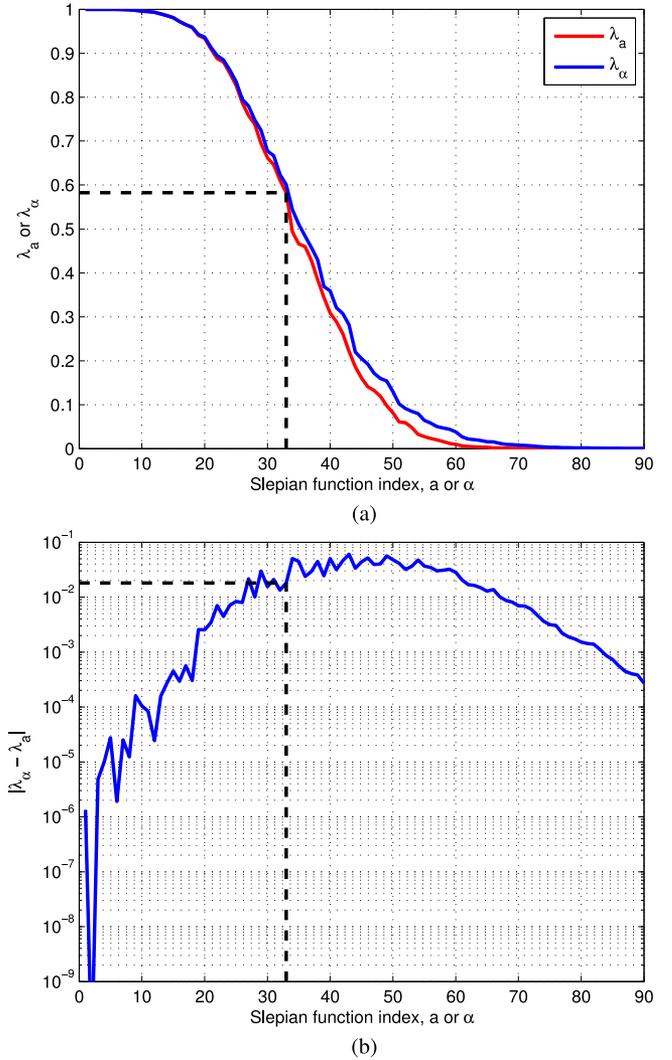

Fig. 9. a) Eigenvalue spectrum for Slepian functions concentrated in South America band-limited at $L = 32$ ($N_\Theta = 90$) obtained using the proposed $\lambda_a$, $a = 1, 2, \ldots, N_\Theta$ and conventional $\lambda_\alpha$, $\alpha = 1, 2, \ldots, N_\Theta$ methods. b) Absolute difference in eigenvalues $|\lambda_\alpha - \lambda_a|$. Dashed back line shows approximate number of well-concentrated Slepian functions $N$.

Slepian functions obtained using the conventional and the proposed method, $10 \log_{10} |h_\alpha - f_a|$, for the ten most concentrated Slepian functions is shown in Fig. 11. The maximum difference observed in Fig. 11 is smaller than 0 dB, as was the case for the example of Slepian functions with $L = 64$ defined on mainland Australia in Fig. 5, showing that the proposed method allows for accurate computation of Slepian functions.

## VI. CONCLUSION

We have proposed a new method for the computation of Slepian functions on the sphere for an arbitrary spatial region. By exploiting the efficient computation of Slepian functions for the polar cap region on the sphere, we have developed a formulation, supported by a fast algorithm, for the approximate computation of Slepian functions for arbitrary spatial region. In comparison to the conventional method of computing Slepian functions, the proposed method enables faster computation and has manageable storage requirements. We derive the approximation error and define the quality of approximation measure as the ratio of energy of the approximation to the energy of the true Slepian function in the region of interest. We have conducted numerical experiments to show that the proposed method maintains accurate computation of Slepian functions and has a high quality of approximation, allows for faster computation and has significantly smaller storage requirements than the conventional method. The proposed method enables accurate computation for Slepian functions for an arbitrary region which does not have to be well-approximated by a rotational symmetric region and is particularly efficient in terms of computational complexity and storage requirements when the region has an enclosing rotationally symmetric region with a small area.

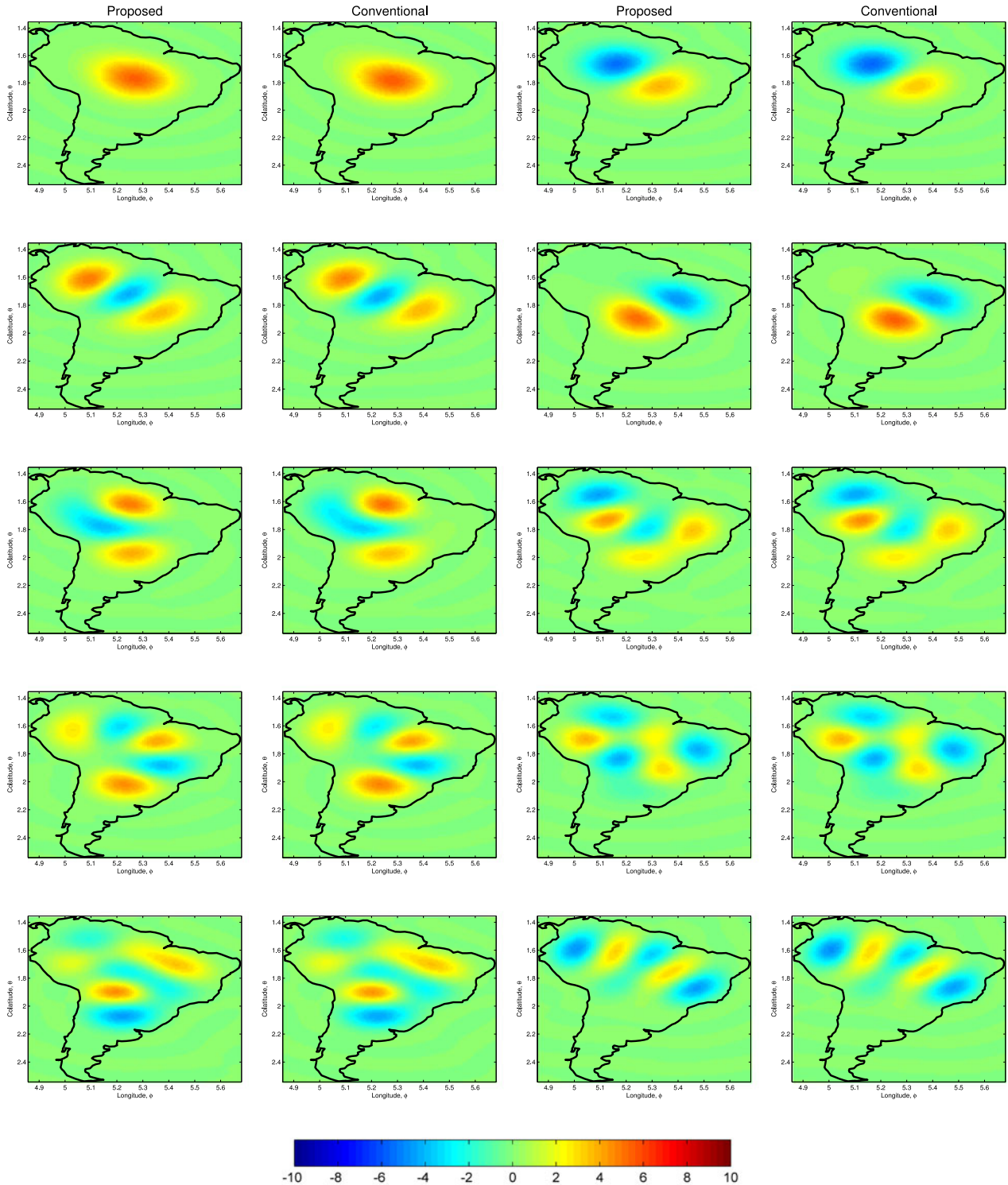

Fig. 10. Slepian functions $f_a(\hat{\boldsymbol{x}})$, $a = 1, 2, \ldots, 10$ and $h_\alpha(\hat{\boldsymbol{x}})$, $\alpha = 1, 2, \ldots, 10$ most concentrated in South America with band-limit $L = 32$. The ordering of concentration is left to right, top to bottom with Slepian functions obtained using the proposed method $f_a(\hat{\boldsymbol{x}})$ in the first and third columns and Slepian functions obtained using the conventional method $h_\alpha(\hat{\boldsymbol{x}})$ in the second and fourth columns.

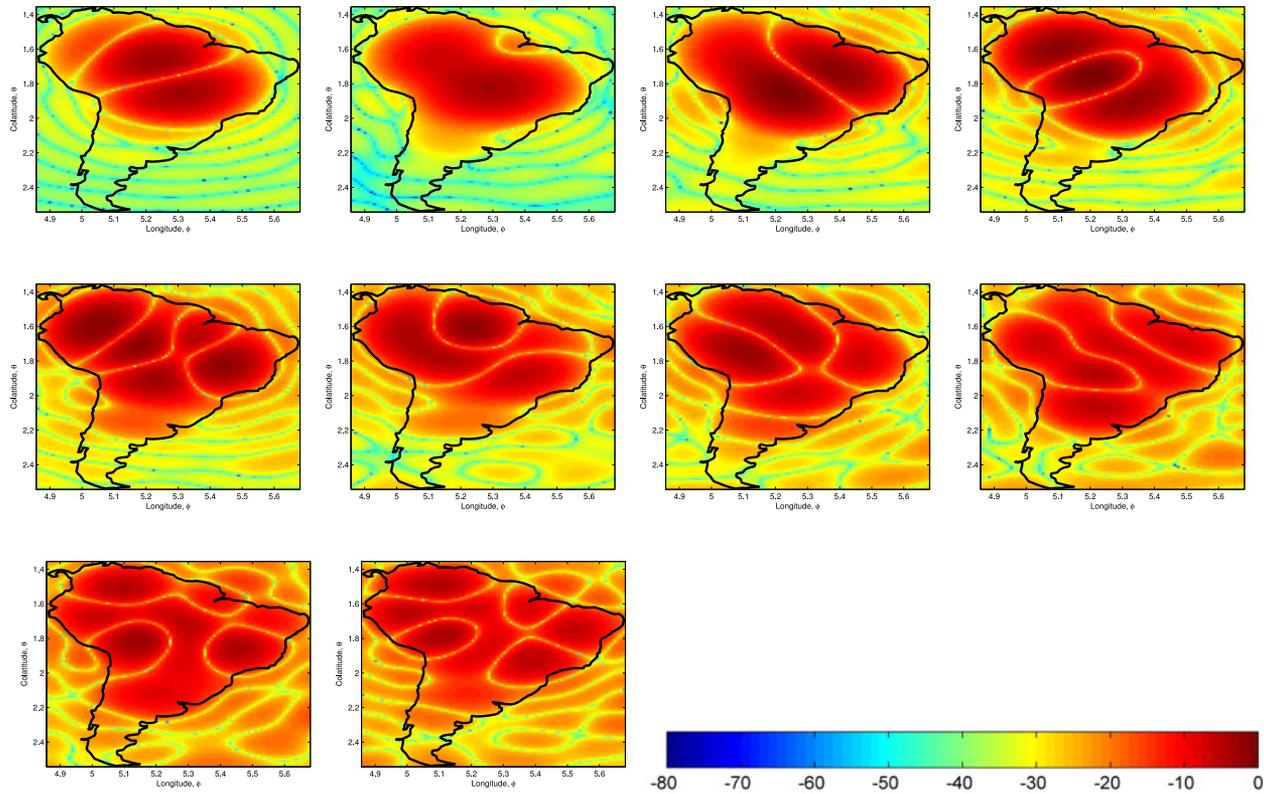

Fig. 11. The difference in decibels between Slepian functions computed using the conventional and the proposed method, $10\log_{10}|h_\alpha - f_a|$, for the ten most concentrated in South America with band-limit $L = 32$. The ordering of concentration is left to right, top to bottom.

The ability to compute Slepian functions with reduced computation time and storage requirements while maintaining accurate computation of Slepian functions will allow for Slepian functions to be used in applications where the data enables large band-limits. Future work includes exploiting the inherently parallel structure of our proposed method to further reduce the time required to compute the Slepian functions. We intend to apply the proposed method to large band-limit applications in a range of fields.

ACKNOWLEDGMENT

The authors would like to thank F. J. Simons for discussions, and the Department of Geosciences at Princeton University for their hospitality during Alice P. Bates' visit in the Fall of 2015.

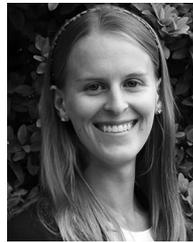

**Alice P. Bates** (S'11–M'16) received the B.E. (first class Hons.) degree in electrical engineering from the University of Auckland, Auckland, New Zealand, in 2013. She received the Ph.D. degree in engineering from the Australian National University (ANU), Canberra, Australia, in September 2016.

She is currently working as a Research Fellow at the Research School of Engineering, ANU. She received the Senior Scholar Award for first in her cohort during her undergraduate studies. She received an Australian Postgraduate Award for the duration of her Ph.D. Her research interests are related to the application driven development of signal processing techniques for the collection and processing of signals with spherical geometry.

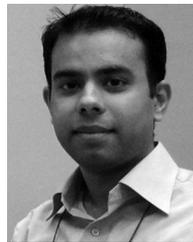

**Zubair Khalid** (S'10–M'13) received the B.Sc. (first class Hons.) degree in electrical engineering from the University of Engineering and Technology, Lahore, Pakistan, in 2008. He received the Ph.D. degree in engineering from the Australian National University of Canberra, Canberra, Australia, in August 2013.

He is currently working as an Assistant Professor in the Department of Electrical Engineering, University of Engineering and Technology, Lahore, Pakistan. Previously, he was working as a Research Fellow in the Research School of Engineering, Australian National University, Canberra, Australia. He was awarded University Gold Medal and Industry Gold Medals from Siemens and Nespak for his overall outstanding performance in Electrical Engineering during his undergraduate studies. He received an Endeavour International Postgraduate Award for his Ph.D. studies. His research interests include the area of signal processing and wireless communications, including the development of novel signal processing techniques for signals on the sphere and the application of stochastic geometry in wireless ad hoc networks.

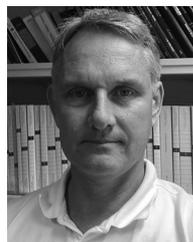

**Rodney A. Kennedy** (S'86–M'88–SM'01–F'05) received the B.E. degree (first class Hons. and university medal) from the University of New South Wales, Sydney, Australia, the M.E. degree from the University of Newcastle, NSW, Australia, and the Ph.D. degree from the Australian National University, Canberra, Australia.

Since 2000, he has been a Professor in engineering at the Australian National University. He has co-authored more than 300 refereed journal or conference papers and a book "Hilbert Space Methods in Signal Processing" (Cambridge Univ. Press, 2013). He has been a Chief Investigator in a number of Australian Research Council Discovery and Linkage Projects. His research interests include digital signal processing, digital and wireless communications, and acoustical signal processing.